\documentclass{pasa}%

\usepackage{graphicx}
\usepackage{hyperref}
\usepackage[table]{xcolor}
\usepackage{graphics,psfrag}
\usepackage{graphicx,psfrag}

\title[Tabulated Neutron Star Equations of State Modeled within the Chiral Mean Field Model]{Tabulated Neutron Star Equations of State Modeled within the Chiral Mean Field Model}

\author[V. Dexheimer]{{V. Dexheimer}
\affil{Department of Physics, Kent State University, Kent OH 44242 USA}%
}%

\jid{PASA}
\doi{10.1017/pas.\the\year.xxx}
\jyear{\the\year}

\usepackage{aas_macros}
\usepackage{hyperref} 
\hypersetup{colorlinks,citecolor=blue,linkcolor=blue,urlcolor=blue}

\hypersetup{draft}

\begin{document}

\begin{frontmatter}
\maketitle

\begin{abstract}
In this special issue article, I review some of the accomplishments of the chiral mean field (CMF) model, which contains nucleon, hyperon, and quark degrees of freedom, and its applications to proto-neutron and neutron stars. I also present a set of equation of state and particle population tables built using the CMF model subject to physical constraints necessary to reproduce different environments, such as those present in cold neutron stars, core-collapse supernova explosions and different stages of compact star mergers.
\end{abstract}

\end{frontmatter}

\section{INTRODUCTION }
\label{sec:intro}

Neutron star interiors cover an incredible range of densities going from about $10^4$ g/cm$^3$ in the crust to about $10^{15}$ g/cm$^3$ in the stellar core, corresponding to several times the nuclear saturation density. Although matter is reasonably understood until a little bit beyond saturation density, unfortunately,  not much is known for larger densities. An exception is matter at extremely high densities, beyond the ones reached in the core of neutron stars, when perturbative QCD (PQCD) provides reliable results (\cite{Freedman:1977gz,Freedman:1976ub,Freedman:1976dm,Freedman:1976xs,Andersen:2002jz,Fraga:2013qra}). Since lattice QCD has not yet been extrapolated to high densities, a natural choice of description for such environments relies on effective models. 

Effective models, after being calibrated to work on a certain regime of energies, can produce reliable results concerning the matter equation of state (EoS) and associated particle population, which in the case of neutron stars, can be further used in dynamical simulations. These simulations include core collapse supernova explosions, star cooling, and compact star mergers and require tabulated data with information about the mycrophysics as input. In this article, I present tables calculated within the chiral mean field (CMF) model that can be used for this purpose.
 
\section{The CMF Model}

Since the same physical laws govern particles under all conditions, it is only logical that a model used to describe cold and dense stars should also be compatible with the description of hot environments, such as the ones created in supernova explosions and compact star mergers, and ultimately in heavy-ion collisions or the early universe. After all, these environments are only different facets of a larger picture of matter with high energy, represented in the QCD phase diagram. Although fully evolved neutron stars have temperatures $\lesssim$1\ MeV, proto-neutron stars can reach temperatures of about 30 MeV or more in their centers (\cite{Burrows:1986me,Pons:1998mm}) and, in compact star mergers, temperatures of approximately 80 MeV can be reached (\cite{Galeazzi:2013mia}). Such temperatures are not significantly different from temperatures reached in high-energy heavy-ion collisions (such as the ones performed in the RHIC and LHC particle colliders (\cite{Schenke:2011tv,Alqahtani:2017jwl}). Having those numbers in mind, it becomes natural to use the same mathematical model, or at least the same kind of approach to describe all such systems.

The CMF model is based on a non-linear realization of the SU(3) sigma model. It is an effective quantum relativistic model that describes hadrons interacting via meson exchange and it is constructed in a chirally-invariant manner, as the particle masses originate from interactions with the medium and, therefore, decrease at high densities/temperatures. The non-linear realization of the sigma model is an improvement over the widely-used sigma model and it includes the pseudoscalar mesons as the angular parameters for the chiral transformation. In this case, the pseudoscalar mesons exhibit a pseudovector coupling to the baryons in agreement with the experimental finding of a vanishing $\pi-N$ scattering length, chiral invariance for heavy particles is ensured if their coupling is invariant under local SU(3) vector transformations (allowing couplings between baryons and meson octets), a connection to the phenomenological Walecka model exists, the masses of the pseudoscalar mesons do not become imaginary at high densities, etc. As a consequence, in addition to reproducing chiral symmetry restoration, the model is in very good agreement with nuclear physics data (\cite{Papazoglou:1998vr}).

The Lagrangian density of the CMF model within the mean field approximation reads
\begin{eqnarray}
\mathcal{L} = \mathcal{L}_{\rm{Kin}}+\mathcal{L}_{\rm{Int}}+\mathcal{L}_{\rm{Self}}+\mathcal{L}_{\rm{SB}}\,,
\end{eqnarray}
where, besides the kinetic energy term for baryons (entire octet) and free leptons (included to ensure charge neutrality), the terms
\begin{eqnarray}
\mathcal{L}_{\rm{Int}}&=&-\sum_i \bar{\psi_i}[\gamma_0(g_{i\omega}\omega+g_{i\phi}\phi+g_{i\rho}\tau_3\rho)+M_i^*]\psi_i,\nonumber\\
\mathcal{L}_{\rm{Self}}&=&+\frac{1}{2}(m_\omega^2\omega^2+m_\rho^2\rho^2+m_\phi^2\phi^2)\nonumber\\
&+&g_4 (\omega^4 + 3 \omega^2 \phi^2 + \frac{\phi^4}{4} + \frac{4 \omega^3 \phi}{\sqrt{2}} + \frac{2 \omega \phi^3}{\sqrt{2}})\nonumber\\
&-&k_0(\sigma^2+\zeta^2+\delta^2)-k_1(\sigma^2+\zeta^2+\delta^2)^2\nonumber\\
&-&k_2\left(\frac{\sigma^4}{2}+\frac{\delta^4}{2}+3\sigma^2\delta^2+\zeta^4\right)-k_3(\sigma^2-\delta^2)\zeta\nonumber\\
&-&k_4\ \ \ln{\frac{(\sigma^2-\delta^2)\zeta}{\sigma_0^2\zeta_0}}\,,\nonumber\\
\mathcal{L}_{\rm{SB}}&=&-m_\pi^2 f_\pi\sigma-\left(\sqrt{2}m_k^ 2f_k-\frac{1}{\sqrt{2}}m_\pi^ 2 f_\pi\right)\zeta\,,
\end{eqnarray}
represent the interactions between baryons and vector and scalar mesons, the self interactions of scalar and vector mesons, and an explicit chiral symmetry breaking term, which is responsible for producing the masses of the pseudo-scalar mesons. The mesons included are the vector-isoscalars $\omega$ and $\phi$ (strange quark-antiquark state), the vector-isovector $\rho$, the scalar-isoscalars $\sigma$ and $\zeta$ (strange quark-antiquark state) and  the scalar-isovector $\delta$, with $\tau_3$ being twice the isospin projection operator of each particle ($\pm$1). The isovector mesons affect isospin-asymmetric matter and thus, are important for neutron star physics. Also, the $\delta$ meson has a contrary but complementary role to the $\rho$ meson, much like the $\sigma$ and $\omega$  mesons. The finite-temperature calculations include a heat bath of hadronic and quark quasiparticles and their antiparticles within the grand canonical potential of the system.

\begin{table}
\caption{Coupling constants for the model, using $\chi = 401.93$ MeV.}
\centering
\begin{tabular}{ccc}
\hline\hline
\\[-2ex]
$ g_{N\omega}=11.90 $&$ g_{N\rho}=4.03 $&$ g_{N\phi}=0 $\\
$ g_{N\sigma}=-9.83 $&$ g_{N\delta}=-2.34 $&$ g_{N\zeta}=1.22$ \\
$ g_{\Lambda\omega}=7.93 $&$ g_{\Lambda\rho}=0 $&$ g_{\Lambda\phi}=-7.32 $ \\
$ g_{\Lambda\sigma}=-5.52 $&$ g_{\Lambda\delta}=0 $&$ g_{\Lambda\zeta}=-2.30$ \\
$ g_{\Sigma\omega}=7.93 $&$ g_{\Sigma\rho}=7.93 $&$ g_{\Sigma\phi}=-7.32 $ \\
$ g_{\Sigma\sigma}=-4.01 $&$ g_{\Sigma\delta}=-6.95 $&$ g_{\Sigma\zeta}=-4.44 $ \\
$ g_{\Xi\omega}=3.97 $&$ g_{\Xi\rho}=3.97 $&$ g_{\Xi\phi}=-14.65 $ \\
$ g_{\Xi\sigma}=-1.67 $&$ g_{\Xi\delta}=-4.61 $&$ g_{\Xi\zeta}=-7.75 $ \\
\\[-2ex]
\hline
\\[-2ex]
$ g_4=38.90 $&$ k_0=1.19 \chi^2 $&$ k_1=-1.40 $ \\
$ k_2=5.55 $&$ k_3=2.65 \chi $&$ k_4=-0.02 \chi^4 $ \\
\\[-2ex]
\hline\hline
\end{tabular}
\label{tabela1}
\end{table}

The effective masses for the baryons are generated by the scalar mesons
\begin{eqnarray}
M_{i}^*&=&g_{i\sigma}\sigma+g_{i\delta}\tau_3\delta+g_{i\zeta}\zeta+M_{0_i}\,,
\end{eqnarray}
with the exception of small explicit mass terms $M_{0_N}=151.68$ and $M_{0_{\Lambda,\Sigma,\Xi}}=354.91$ MeV.

The coupling constants of the model were presented in Ref.~\cite{Dexheimer:2008ax} and are shown here in Table~1. The scalar sector was fitted to reproduce the vacuum masses of baryons and mesons and the pion and kaon decay constants. The vector sector was fitted to reproduce nuclear constraints for symmetric matter at saturation, such as baryon density ($\rho_0=0.15$ fm${-3}$), binding energy per nucleon ($B/A=-16$ MeV), compressibility ($K=300$ MeV), as well as symmetry energy ($E_{\rm{sym}}=30$ MeV), and symmetry energy slope ($L=88$ MeV). The reproduced pressure and compressibility for neutron matter at saturation are $P=4.56$ MeV/fm$^3$ and $K=870$ MeV. The reproduced hyperon potentials at saturation are $U_\Lambda=-28$ MeV, $U_\Sigma=5$ MeV, $U_\Xi=-18$ MeV. The reproduced critical point for the nuclear liquid-gas phase transition lies at $T_c=16.4$ MeV, ${n_B}_c=0.05$ fm$^{-3}$, ${\mu_B}_c=910$ MeV.

Regarding the constraints for the slope of the symmetry energy $L$, although
a compilation of several studies indicates values lower than 60 MeV (\cite{Lattimer:2012xj}),
there are other works which indicate that the values of such quantity should be larger than 90 MeV
(\cite{Cozma:2013sja,Chen:2005ti,Sotani:2015lya,Tsang:2012se,Wang:2014rva}). It is also important to note that the values suggested in Ref.~\cite{Lattimer:2012xj} are a result of a compilation of experimental analyses that have different systematic and statistical errors, hence, should be interpreted  carefully when it comes to excluding equations of state.

The numerical code for the CMF model solves a set of equations for each baryon chemical potential and temperature. Those include an equation of motion for each meson and one extra equation in the case that baryon number density is fixed (instead of chemical potential). Additional constraints such as, for example, charge neutrality, fixed charge fraction, fixed lepton fraction, fixed entropy per baryon and zero net isospin require additional equations.

In order to study neutron stars, charge neutrality and chemical equilibrium are required. As a result of the energy balance, highly isospin-asymmetric objects are formed. Nevertheless, in proto-neutron stars, the proton-to-neutron ratio is larger than in cold neutron stars, making these systems more similar to heavy-ion collision environments. Studies of the CMF model including fixed entropy per baryon together with trapped neutrinos were able to reproduce massive neutron stars (like the ones that have been recently observed (\cite{Antoniadis:2013pzd,Demorest:2010bx})) even taking hyperons into account (\cite{Dexheimer:2008ax}). The CMF formalism was also used to study the effect of kaon condensation in neutron and proto-neutron stars (\cite{Mishra:2009bp}), the inclusion of chiral partners in stars (\cite{Dexheimer:2007tn,Dexheimer:2008cv,Dexheimer:2014pea,Mukherjee:2017jzi}), and the cooling profile of stars (\cite{Negreiros:2010hk,Dexheimer:2015qha}), which was (and still is) in good agreement with observed data. For cold chemically-equilibrated stars, a maximum mass star with $2.1$ M$_{\odot}$ and corresponding radius of $12$ km is reproduced ($1.93$ M$_{\odot}$ and $13$ km when quarks are included). For the canonical star with mass $1.4$ M$_{\odot}$, a corresponding radius of $14$ km is found.

Up, down, and strange quarks were introduced in the formalism within the same model (\cite{Dexheimer:2009hi,Negreiros:2010hk,Hempel:2013tfa}), however, the degrees of freedom which are actually populated at a certain temperature and density change from hadrons to quarks and vice-versa through the introduction of an extra field $\Phi$ in the effective masses of baryons and quarks (Eqs.~6 and 7 in Ref.~\cite{Dexheimer:2008ax}). The scalar field $\Phi$ was named in analogy with the Polyakov loop (\cite{Fukushima:2003fw}), since it also functions as the order parameter for deconfinement. The potential for $\Phi$ (see Eq.~(9) in Ref.~(\cite{Dexheimer:2008ax}))  was modified from its original form in the Polyakov-loop-extended Nambu-Jona-Lasinio (PNJL) model (\cite{Ratti:2005jh,Roessner:2006xn})  in order to be used to study low-temperature and high-density environments (in addition to high-temperature and low-density environments). It was shown in Refs. \cite{Fukushima:2010pp,Lourenco:2012dx,Lourenco:2012yv,Blaschke:2013rma} that this choice for the potential $U(\Phi,T,\mu_B)$ can also be used in the Polyakov-Nambu-Jona-Lasinio model, successfully reproducing QCD features. Nevertheless, the CMF model is significantly different from the widely-used PNJL model (\cite{Fukushima:2003fw}). In our case, both hadronic and quark (in addition to leptonic) degrees of freedom are included. Note that, although a relation between the effective mass for the fermions and the chiral condensate can be written using mean field theory in both models, such a derivation is quite different in each case.

\begin{figure}
\begin{center}
\includegraphics[trim={1.1cm 0 0 2.cm},clip,width=9.5cm]{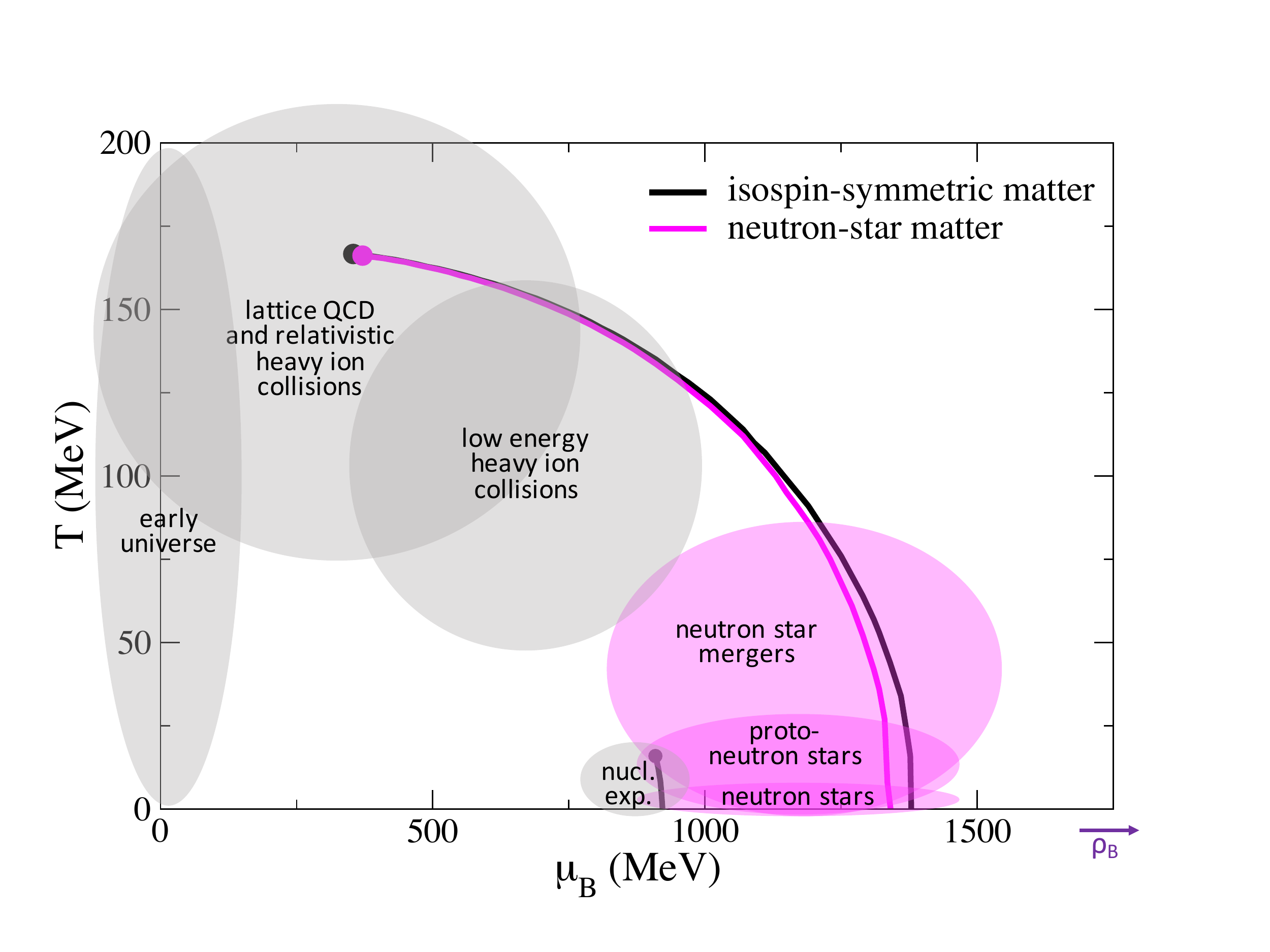}
\caption{QCD Phase diagram resulting from the CMF model. The lines represent first-order transitions. The circles mark the critical end-points. Isospin-symmetric matter refers to zero isospin and strangeness constraints, while neutron-star matter stands for charged neutral matter in chemical equilibrium. The shaded regions exemplify some of the different regimes that can be described within the model.
}\label{Fig1}
\end{center}
\end{figure}
  
The QCD phase diagram shown in Fig.~1 was constructed by analyzing the behavior of the deconfinement to quark matter (and associated chiral symmetry restoration) at different temperatures and chemical potentials (or densities) using the formalism described above for isospin-symmetric matter (zero isospin and strangeness constraints) and neutron-star matter (charge neutral and in chemical equilibrium). The lines represent first-order phase transitions, meaning that for each baryon chemical potential and temperature there are multiple metastable solutions in order-parameter space, although only one truly stable phase exists. The circles mark critical points, beyond which the deconfinement and chiral transitions become smooth crossovers. The shaded regions exemplify some of the different regimes that can be described within the CMF model (that can be applied to the entire $\mu_B-T$ plane). Note that at low density/temperature, the nuclear physics liquid-gas phase transition is also reproduced.

Since the coupling constants related to quark matter cannot be related to nuclear properties, they were constrained using lattice QCD data as well as information about the QCD phase diagram from Refs.~\cite{Ratti:2005jh,Roessner:2006xn,Aoki:2006we,Fodor:2004nz} for symmetric matter (See Table~2 in Ref.~\cite{Dexheimer:2009hi}). In this approach, the chiral symmetry restoration and deconfinement phase transitions range from sharp first-order phase transitions to smooth crossovers as the temperature increases. With the advent of RHIC and LHC, relativistic heavy-ion collision experiments have focused on the low-$\mu_B$ part of the QCD phase diagram. The RHIC Beam Energy Scan is currently investigating the central region of the diagram, and future facilities, like FAIR in GSI and NICA in Dubna, will be capable of a more in-depth exploration of the high-$\mu_B$ region. Lattice QCD calculations are also slowly advancing towards the middle of the diagram (from the left). Nevertheless, only neutron stars will be able to probe the very right side of the QCD phase diagram due to their incredibly high chemical potential to temperature ratio.

The most important and unique aspect of our description is that hadrons are included as quasi-particle degrees of freedom in a chemically-equilibrated mixture with quarks. Therefore, the model gives a quasi-chemical representation of the deconfinement phase transition (so-called chemical picture in terms of electromagnetic non-ideal plasmas (\cite{iosilevskiy2000})). The assumed full miscibility of hadrons and quarks is, for example, in contrast to the underlying picture of simple quark-bag models. At sufficiently high temperature, this will lead to the appearance of quarks in the hadronic sea. On the other hand, it is also possible that some hadrons survive in the quark sea. Nevertheless, quarks will always give the dominant contribution in the quark phase, and hadrons in the hadronic phase. The hadronic and the quark phases are characterized and distinguished from each other by the values of the order parameters, $\sigma$ and $\Phi$. The inter-penetration of quarks and hadrons in the two phases is physical, and is required to obtain a true crossover transition at low baryon chemical potential, which has been shown to exist by lattice QCD calculations.

It is important to note at this point, that astrophysical EoS's including quark deconfinement are usually only carried out up to a few tens of MeV's, if they include finite-temperature effects at all. Different EoS's (representing different degrees of freedom) are put together ``by hand'' in different regions of the neutron star, generating necessarily first-order phase transitions between them. This approach is not compatible with QCD calculations such as the one in Refs.~\cite{Baym:2008me,Lourenco:2012yv,Kojo:2014rca,Masuda:2015kha}, which suggests that the phase transition to quark matter might be a crossover, even at low temperatures. Unlike this approach, our description predicts different degrees of freedom appearing self-consistently at different densities and temperatures. This allowed for the first detailed comparison between the nature of the deconfinement phase transition and the one of the nuclear matter liquid-gas phase transition (\cite{Hempel:2013tfa}). It was found in this work (in agreement with Ref.~\cite{Bombaci:2009jt}) that the principle difference between both phase transitions is that, in contrast to the ordinary Van-der-Waals-like phase transition, the phase coexistence line of the deconfinement phase transition has a negative slope in the pressure-temperature plane. This feature is related to the quark phase having higher entropy per baryon than the hadronic phase. As another qualitative difference, we found that the non-congruent features of the deconfinement phase transition become vanishingly small around the critical point. Non-congruent phase transitions occur for first-order phase transitions with more than one globally conserved charge, allowing local concentrations of the charges to vary during a phase transition, i.e., the appearance of mixtures of phases.

As a final test of the validity of the CMF model, I present here for the first time a figure showing the speed of sound reproduced by the model when a mixture of hadronic and quark phases is allowed. Naturally, as in any relativistic formalism, the speed of sound does not go above 1 (the value of the speed of light in natural units), but note that neither does our speed of sound go above the speed of sound limit provided by kinetic theory (\cite{Moustakidis:2016sab}) given by
\begin{eqnarray}
v_s&=&\sqrt{\frac{\varepsilon-P/3}{P+\varepsilon}}\,.
\end{eqnarray}
Finally, it is important to note that the equation of state presented in Fig.~2 (black line) has already been successfully compared with PQCD calculations in Ref.~\cite{Dexheimer:2017pom}. Other models commonly used to describe quark matter in stars, such as the bag model, do not agree with PQCD calculations (Ref.~ \cite{Fraga:2013qra}).

\begin{figure}
\begin{center}
\includegraphics[trim={1.5cm 0 0 2.8cm},clip,width=9.cm]{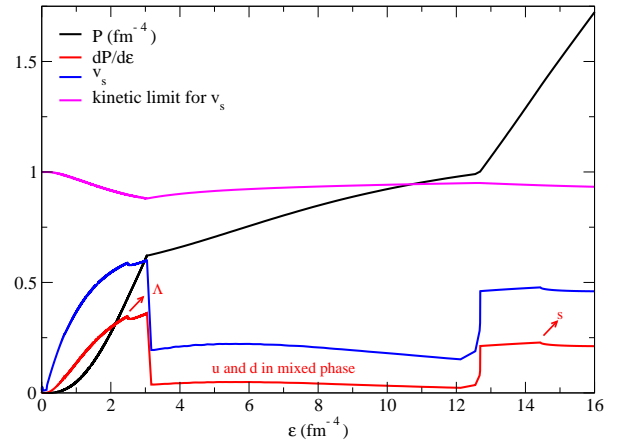}
\caption{Equation of state for neutron-star matter at zero temperature within the CMF model, its derivative and square root of derivative (speed of sound). The kinetic limit for the speed of sound from Ref.~\cite{Moustakidis:2016sab} is also shown. The arrows mark the threshold for the appearance of the Lambdas and strange quarks.
}\label{Fig2}
\end{center}
\end{figure}

\section{Tabulated EoS}

The tables described in this section contain the equation of state (a large set of thermodynamical quantities) and particle population for different sets of conditions calculated using the CMF model. These tables are available in two locations, in the CompOSE website \url{http://compose.obspm.fr/} and in the following website \url{http://personal.kent.edu/~vdexheim/}. At this point in time, there are three different kinds of tables available produced using the CMF model, all containing only hadronic degrees of freedom (and in some cases leptons). In the future, equivalent tables will be uploaded that also contain quark degrees of freedom. The tables, produced under the assumption of different conditions, are
\begin{itemize}
\item 1D tables for chemically-equilibrated neutron stars at zero temperature. In this case, the baryon number density ranges from $0.03$ to $3.03$ fm$^{-3}$ with steps of $0.01$ fm$^{-3}$. They contain contributions from nucleons, hyperons, electrons, and muons.
\item 3D tables for supernova/neutron-star merger simulations without leptons. In this case, the charge fraction ranges from $0$ to $0.53$ with steps of $0.01$, the temperature ranges from $0$ to $160$ MeV with steps of $2$ MeV, and the baryon number density ranges from $0.01$ to $3.01$ fm$^{-3}$ with steps of $0.01$ fm$^{-3}$. They contain contributions from nucleons and hyperons.
\item 3D tables for supernova/neutron-star merger simulations with electrons. In this case, the charge fraction ranges from $0$ to $0.53$ with steps of $0.01$, the temperature ranges from $0$ to $160$ MeV with steps of $2$ MeV, and the baryon number density ranges from $0.01$ to $3.01$ fm$^{-3}$ with steps of $0.01$ fm$^{-3}$. They contain contributions from nucleons, hyperons, and electrons.
\end{itemize}

For each of those sets of conditions, five tables are provided. They are CompOSE standard data files (more information about them can be found in the following website \url{http://compose.obspm.fr/manual/}) and include
\begin{itemize}
\item {\bf{eos.thermo}}: a table with 2 dimensions, 1 316 574 grid points, and 1 316 575 lines (the first line gives the masses of the neutron and proton). The standard columns contain the pressure divided by the baryon number density (in MeV), entropy density divided by the baryon number density, scaled and shifted baryon chemical potential, scaled charge chemical potential, scaled effective lepton chemical potential, scaled free energy density divided by the baryon number density, and scaled internal energy density divided by the baryon number density. Scaled means divided by the nucleon mass, and shifted means it had 1 subtracted.
The two extra columns contain the scaled enthalpy density per baryon number density and strangeness number density ($\sum_i s_i n_i$ in fm$^{-3}$). For the tables containing a quark phase, a third extra column containing the quantity $\sum_i  {Q_B}_i n_i$ used to calculate the total hadronic electric charge will be added (see discussion of Eqs.~(7) and (11)).
\item {\bf{eos.compo}}: a table with 2 dimensions, 1 316 574 grid points, and 1 316 574 lines. The standard columns contain the particle fractions ($n_i/n_B$) and the index encoding the type of phase, which in our case is 1 for hadronic matter and 2 for quark matter.
\item {\bf{eos.nb}}: a table with 1 dimension, 301baryon number density grid values (in fm$^{-3}$), and 303 lines (the first two lines contain the initial and final grid points).
\item {\bf{eos.t}}: a table with 1 dimension, 81temperature grid values (in MeV), and 83 lines (the first two lines contain the initial and final grid points). 
\item {\bf{eos.yq}}: a table with 1 dimension, 54 charge fraction grid values, and 56 lines (the first two lines contain the initial and final grid points).
\end{itemize}

In the case of the 1-dimensional tables for chemically-equilibrated neutron stars at zero temperature, charge neutrality is fulfilled
\begin{eqnarray}
\sum_i {Q_e}_i n_i=0\ ,
\end{eqnarray}
where ${Q_e}_i$ is the electric charge and $n_i$ the number density of each baryon, lepton, or quark and chemical equilibrium is ensured through 
\begin{eqnarray}
\mu_i={Q_B}_i \mu_B+ {Q_e}_i \mu_q\ ,
\label{mu}
\end{eqnarray}
where $\mu_i$ is the chemical potential of each baryon or quark, ${Q_B}_i$ is the baryon number, ${\mu_B}$ the baryon chemical potential, and  $\mu_q$ the charged chemical potential, which is equal to minus the electron/muon chemical potentials.

In the case of the 3-dimensional tables for supernova/neutron-star merger simulations without leptons, charge neutrality is not required and the leptons do not participate in chemical equilibrium, although Eq.~(\ref{mu}) still holds and $\mu_q$ is only equal to zero in isospin-symmetric matter (as Coulomb interactions are not taken into account). The charge fraction is defined as the total charge over baryon number
\begin{eqnarray}
Y_q=\frac Q B=\frac {\sum_i Q_{e_i} n_i}{\sum_i  {Q_B}_i n_i}\ ,
\label{yc}
\end{eqnarray}
where $\sum {Q_B}_i n_i$ is not the same as baryon number density ${n_B}$, as the latter comes from the derivate of the pressure with respect to the baryon chemical potential and, therefore, also contains a contribution from the the potential $U$ for $\Phi$ (when quarks are included). The second law of thermodynamics can be written for the baryons and quarks as
\begin{eqnarray}
\sum_i \varepsilon_{i}=-\sum_i P_i+T\sum_i s_i+\sum_i \mu_i n_i\ ,
\end{eqnarray}
where $\varepsilon_i$, $P_i$, and $s_i$ are the energy density, pressure, and entropy density of each baryon or quark and $T$ is the temperature. This expression can be rewritten using Eq.~(\ref{mu}) and the definition of $Y_q$
\begin{equation}
\sum_i \varepsilon_{i}=-\sum_i P_i+T\sum_i s_i+\sum_i ({Q_B}_i \mu_B+ {Q_e}_i \mu_q) n_i \ , 
\end{equation}
\begin{eqnarray}
\sum \varepsilon_i=-\sum_i P_i+T\sum_i s_i+\widetilde{\mu}\sum_i  {Q_B}_i n_i \, ,
\end{eqnarray}
where $\widetilde{\mu}$ is the free energy of the system defined as $\widetilde{\mu}=\mu_B+ \mu_q Y_q$. For more details on this derivation, see Ref.~\cite{Hempel:2013tfa}. Note that the total energy density, pressure, and entropy density of the system also contain mesonic and $\Phi$ contributions.

In the case of the 3-dimensional tables for supernova/neutron-star merger simulations with electrons, the electrons still do not participate in chemical equilibrium, although, as before, Eq.~(\ref{mu}) still holds for the baryons and quarks and $\mu_q$ is only equal to zero for isospin symmetric matter. The electron density is determined in order to establish charge neutrality, so from Eq.~(\ref{yc}) one can derive
\begin{eqnarray}
\sum_i {Q_e}_i n_i={\sum_i  {Q_B}_i n_i}Y_q=n_e\ ,
\end{eqnarray}
where $n_e$ is the number density of electrons.

Note that there are no nuclei included in our calculations within the CMF model, as its current parametrization describes only bulk matter. A version of the CMF model, which can describe hot and dense matter but also includes nuclei (as in Refs.~\cite{Papazoglou:1998vr,Beckmann:2001bu,Schramm:2002xi,Schramm:2015bia}) is under way and will be reported in the near future. For this reason, a star crust should be added to the tables presented in this article before using them to perform realistic star simulations.

\begin{figure}
\begin{center}
\includegraphics[trim={0cm 0 0 2.8cm},clip,width=9.cm]{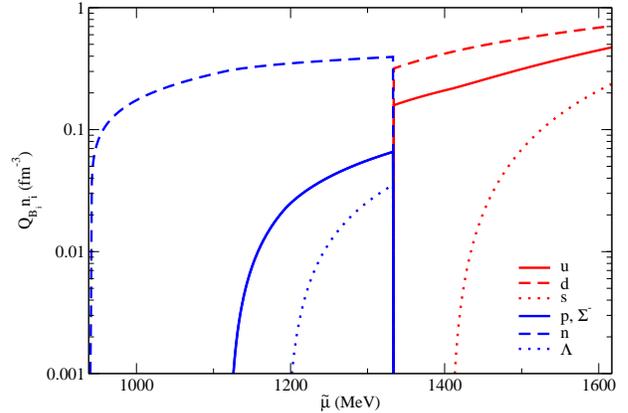}
\caption{Particle population for charge-neutral matter without leptons ($Y_q=0$) at zero temperature within the CMF model.}\label{Fig3}
\end{center}
\end{figure}

\begin{figure}
\begin{center}
\includegraphics[trim={0cm 0 0 2.8cm},clip,width=9.cm]{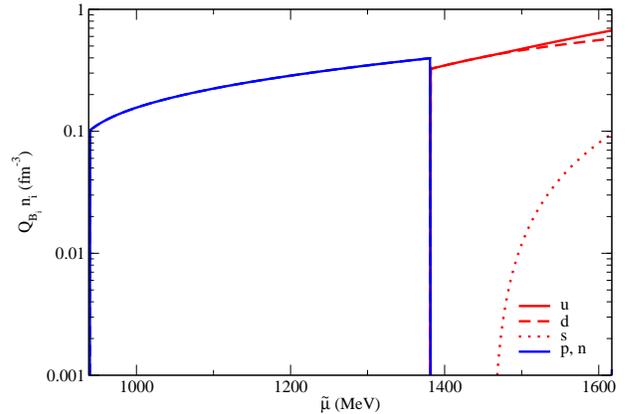}
\caption{Particle population for half-charged matter without leptons ($Y_q=0.5$) at zero temperature within the CMF model.}\label{Fig4}
\end{center}
\end{figure}

As an example of our results containing quarks (without the assumption of mixtures of phases), Figs.~\ref{Fig3} and \ref{Fig4} show the particle population for the extreme cases of charge-neutral mater ($Y_q=0$) and half-charged matter ($Y_q=0.5$) at zero temperature without the inclusion of leptons. In both figures, the y-axis contains the baryon number density, meaning that the quark densities were multiplied by 1/3. In Fig.~\ref{Fig3}, one can see that all hyperons but the Lambdas are suppressed by the appearance of the quarks.  In Fig.~\ref{Fig4}, one can see that all hyperons are suppressed and that the quarks appear at larger free energy $\widetilde{\mu}$. 

In Fig.~\ref{Fig3}, the free energy equals the baryon chemical potential in each phase $\widetilde{\mu}=\mu_{B_H}=\mu_{B_Q}$ but, in Fig.~\ref{Fig4}, the free energy is different from the baryon chemical potential in each phase $\widetilde{\mu}=\mu_{B_H}+ 0.5 \mu_{q_H}=\mu_{B_Q}+ 0.5 \mu_{q_Q}$. For more details on the correspondence of different chemical potentials in different phases, see Ref.~\cite{Hempel:2013tfa}. Finally, note that fixed charge fraction $Y_q=0.5$ does not correspond to isospin symmetric matter if net strangeness is not set to zero, which is the case here. This can easily be verified by the presence of strange quarks in Fig.~\ref{Fig4}.

\section{CONCLUSION}

I presented in this article a review of some of the accomplishments of the CMF model in describing dense and/or hot matter and, in particular, neutron stars under different stages of evolution. These accomplishments also include a possible description of matter produced in heavy-ion collisions (\cite{Steinheimer:2009nn}). 

For the first time a tabulated version of the EoS and particle population produced within the CMF model under different conditions was presented. These tables can be used in simulations of, for example, core-collapse stellar explosion, stellar cooling, and mergers.

Although the CMF tables that are already available online only contain hadronic matter (a version with quarks will be available in the near future), their results contain essential features of the description of hot and/or dense matter, such as chiral symmetry restoration,  the inclusion of hyperons, and the inclusion of antiparticles for all fermions. In addition, since the CMF model results derive from a relativistic description, a subluminal speed of sound is guaranteed in any regime.

\begin{acknowledgements}
The author would like to thank Matthias Hempel for inumerous helpful discussions. The author acknowledges the support from NewCompStar COST Action MP1304.
\end{acknowledgements}

\bibliographystyle{pasa-mnras}
\bibliography{1r_lamboo_notes}

\end{document}